\documentclass[reprint,prb,twocolumn,showpacs,superscriptaddress,aps,longbibliography]{revtex4-2}

\usepackage[colorlinks=true,citecolor=blue]{hyperref} 
\usepackage{xurl} 

\usepackage{xcolor}
\usepackage{graphicx}
\usepackage{amsmath, amssymb, mathrsfs}
\usepackage{dcolumn}
\usepackage{bm}
\usepackage[export]{adjustbox}

\begin{document}

\title{Hamiltonian learning quantum magnets with dynamical impurity tomography}

\author{Netta Karjalainen}
\affiliation{Department of Chemistry, University of Helsinki, Finland}
\affiliation{Department of Applied Physics, Aalto University, 02150 Espoo, Finland}

\author{Greta Lupi}
\affiliation{Department of Applied Physics, Aalto University, 02150 Espoo, Finland}

\author{Rouven Koch}
\affiliation{QuTech and Kavli Institute of Nanoscience, Delft University of Technology, 2628 CJ Delft, The Netherlands}

\author{Adolfo O. Fumega}
\affiliation{Department of Applied Physics, Aalto University, 02150 Espoo, Finland}

\author{Jose L. Lado}
\affiliation{Department of Applied Physics, Aalto University, 02150 Espoo, Finland}

\date{\today}

\begin{abstract}
Nanoscale engineered spin systems, ranging from spins on surfaces to nanographenes, provide flexible platforms to realize entangled quantum magnets from a bottom-up approach. However, assessing the quantum many-body Hamiltonian realized in a specific experiment remains an exceptional open challenge, due to the difficulty of disentangling competing terms accounting for the many-body excitations.  Here, we demonstrate a machine learning strategy to learn a quantum many-body spin Hamiltonian from scanning spectroscopy measurements of spin excitations. Our methodology leverages the spatially resolved reconstruction of the many-body excitations induced by depositing quantum impurities next to the quantum magnet. We demonstrate that our algorithm allows us to predict long-range Heisenberg exchange interactions, anisotropic exchange, and antisymmetric Dzyaloshinskii-Moriya interaction, including in the presence of sizable noise. Our methodology establishes defect-induced spatially resolved dynamical excitations in quantum magnets as a powerful strategy to understand the nature of quantum spin many-body models.  
\end{abstract} \maketitle

\section{Introduction}
Artificial quantum magnets provide a versatile platform to explore quantum
many-body phenomena, including  emergent entangled quantum spin liquid
phases~\cite{Savary2016, Broholm2020, Wang2024}, and enabling
atomic-scale quantum technologies\cite{Wang2024, Hirohata2020, Fukami2024,
Wang2023}.  Manipulation and measurements with scanning probe
microscopy\cite{Eigler1990, Yang2019} provide precise control over microscopic
structure, enabling the exploration of quantum magnets from a bottom-up
strategy~\cite{Spinelli2014, Drost2023}.  Inelastic
spectroscopy\cite{Heinrich2004} and spin resonance\cite{Baumann2015} with
scanning tunneling spectroscopy allow to locally measure spin excitations with
high spatial and energy resolution~\cite{Yang2019, Baumann2015, Paul2016,
Phark2023, Wang2023, Lado2017,Choi2019,Sun2025,
Wang2024,Zhao2024, Lutz2017,Kawaguchi2022}.  Nanoscale magnetic
systems exhibit rich and anisotropic interactions, including
Dzyaloshinskii-Moriya couplings, single-ion anisotropy, and long-range
exchange~\cite{Choi2019}.  When engineering these artificial systems, a crucial
open question is how the Hamiltonian of the system can be precisely obtained
from spectroscopic measurements, especially in cases with multiple competing
interactions.

\begin{figure}[t!]
\center
\includegraphics[width=\linewidth]{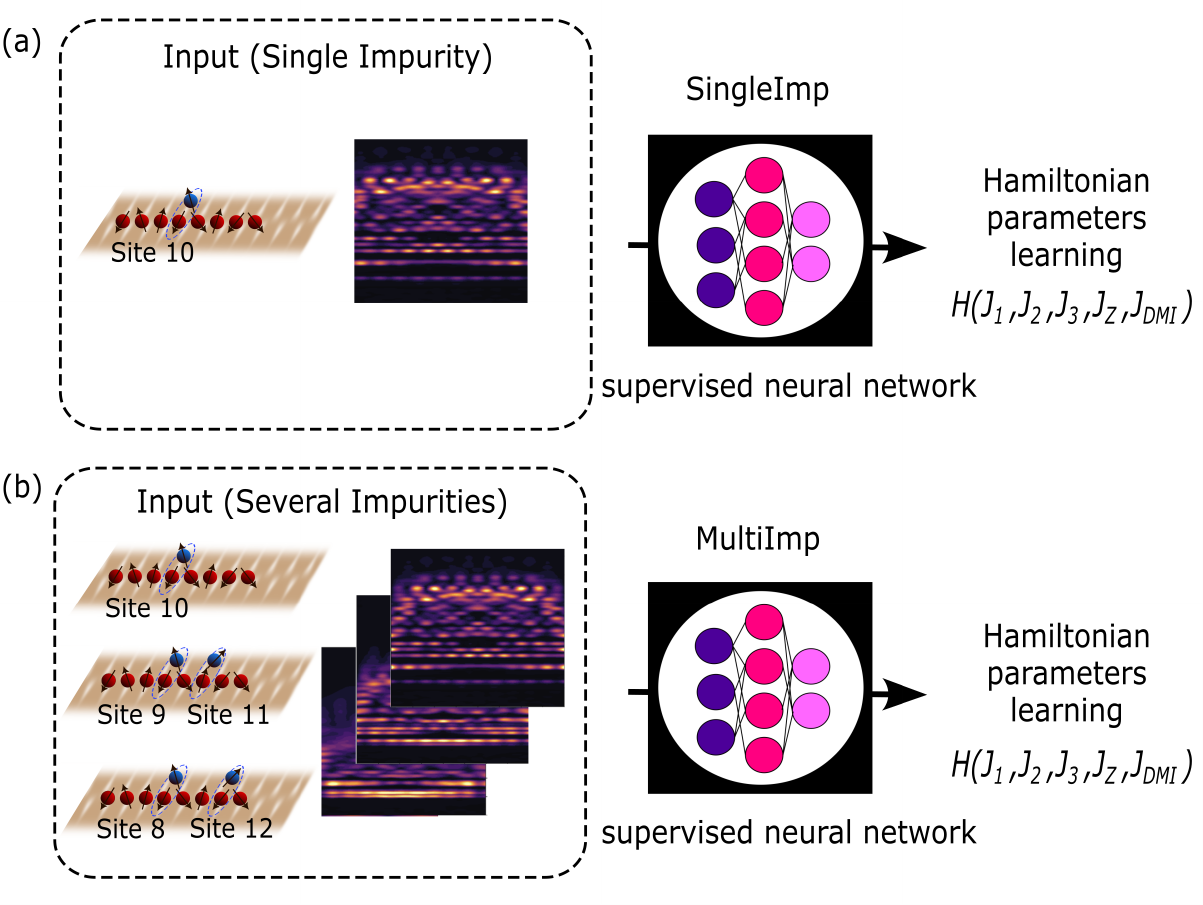}
\caption{\textbf{Hamiltonian learning with impurity tomography}. Two strategies are shown, (a) SingleImp that uses only single impurity placement, and (b) MultiImp that uses several impurity configurations, both single and several impurities, with variable distance between impurities. Dynamical correlators of the spin chain are computed or measured and passed to a machine learning model. The trained neural network then reconstructs the corresponding Hamiltonian of the spin chain.}
\label{Fig:SingleImp and MultiImp}
\end{figure}

Impurities act as local perturbations in a quantum material, revealing
subtle effects in the underlying structure of the ground state and
its many-body excitations.
Reconstruction and scattering
around impurities~\cite{Friedel1952} enable imaging electronic structures
through quasiparticle interference~\cite{Crommie1993,Rutter2007,Avraham2018}, 
and probing their internal geometric topological structure~\cite{Guan2024,Roushan2009,PhysRevLett.104.016401,Zheng2018,Dutreix2019}.
Impurity scattering effects can also be exploited in purely spin systems,
enabling probing quasiparticle interference of quantum
many-body excitations, including
spinons~\cite{PhysRevResearch.2.033466,Chen2022NP,PhysRevB.105.195156,Takahashi2025,PhysRevLett.125.267206,Ruan2021}, 
magnons~\cite{Ganguli2023,PhysRevLett.133.046503,PhysRevX.13.021016,PhysRevLett.130.066701}
and triplons~\cite{Drost2023,Koch2025,Zhao2024}.
However,
extracting complex Hamiltonians by
leveraging impurities remains an open challenge,
as impurity reconstructions become highly complex
in systems with competing interactions.
Machine learning techniques enable new strategies
to characterize correlated and topological
quantum matter~\cite{vanNieuwenburg2017,Carrasquilla2017,RodriguezNieva2019,PhysRevB.102.054107,PhysRevLett.124.226401,PhysRevA.103.012419,Kming2021,PhysRevB.97.115453,Dunjko2018,Torlai2018,PhysRevE.95.062122,Zhang2023,Melnikov2018,PhysRevLett.120.066401,PhysRevA.107.010101,Carrasquilla2020},
including learning Hamiltonians from complex observables~\cite{Karjalainen2023, Valenti2022,f58h-zxs3,PhysRevResearch.4.033223,Tucker2024, samarakoon2022, heightman2024, PhysRevApplied.20.044081,Koch2025,PhysRevB.111.054404,Lupi2026}.
As a result, the combination of machine learning and
local impurities offers a potential strategy to perform
Hamiltonian learning in complex quantum materials~\cite{Khosravian2024,PhysRevB.109.195125,Liu2025,PhysRevB.111.014501,PhysRevResearch.4.043224,Sobral2023,lupi2024,Nigmatulin2026}.

\begin{figure*}[ht!]
\center
\includegraphics[width=\linewidth]{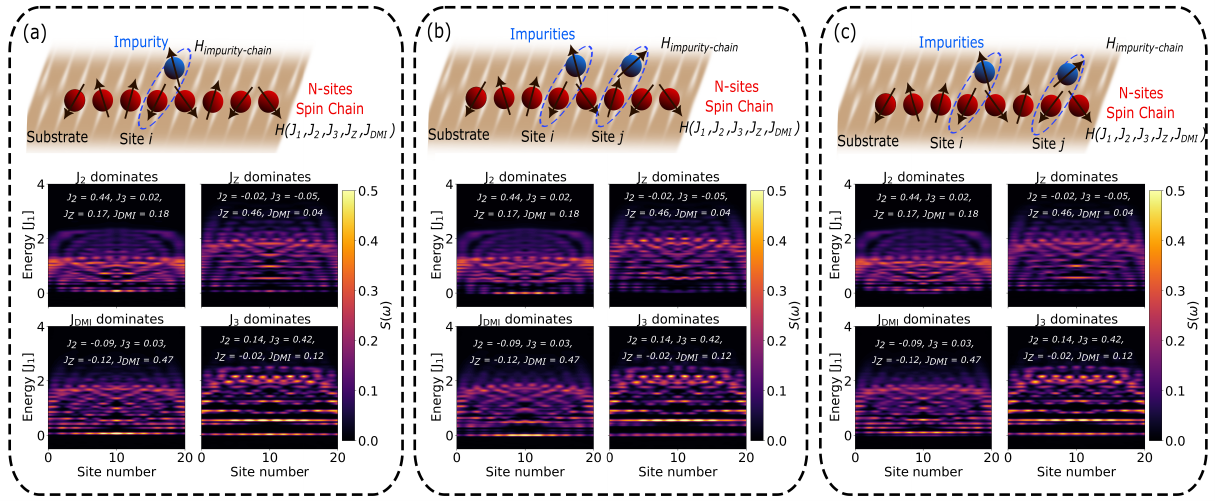}
\caption{\textbf{Impurity configurations and impact in the many-body excitations}. 
Impurities are added to a quantum spin model in different locations, triggering different
many-body reconstructions depending on the Hamiltonian. Examples show how different dominating parameters affect the appearance of the dynamical correlators. (a) Only one impurity is placed next to the spin chain.
(b) Two impurities are placed one spin apart. (c) Two impurities are placed three spins apart.}
\label{Fig:schematic}
\end{figure*}

Here, we show that spin Hamiltonians with competing interactions can be extracted from spatially resolved and frequency-resolved
spin excitations as directly accessible with scanning tunneling microscope (STM) spectroscopy. 
Our methodology relies on a machine learning strategy to extract the Hamiltonian parameters from local spin excitations, which leverages engineered quantum magnetic impurities placed next to spin chains. These impurities trigger a many-body reconstruction of the ground state and excitations of the quantum magnet, 
providing the required
information to reconstruct the underlying unperturbed Hamiltonian. The parameter extraction is robust to the noise in spin excitations and provides the parameters instantly once the algorithm is trained. We show that in the presence of noise, 
providing multiple impurity configurations simultaneously
leads to a more robust
Hamiltonian learning method.
Our manuscript is organized as follows, section \ref{section: quantum spin chains} gives details of the many-body spin model in question;  section \ref{section: hamiltonian learning} describes the machine learning methods used for Hamiltonian parameter extraction, data generation, and the inclusion of noise to simulate experimental conditions; in Section \ref{section: discussion}, we present the results for the Hamiltonian inference; and Section \ref{section: conclusions} summarizes our conclusions.

\section{Model} 
\label{section: quantum spin chains}
Artificial quantum systems offer a versatile platform for engineering desired quantum properties. Here, we focus on spin-$1/2$ Hamiltonian,
as realized on multiple platforms, including
Ti and Cu atoms on MgO\cite{Willke2018,Wang2024,Yang2018}. 
The Hamiltonian of the system takes the form:
\begin{equation}
\label{eqn: hamiltonian}
\begin{split}
    H &=  J_1 \sum_{\langle i,j \rangle} \mathbf{S}_i \cdot \mathbf{S}_j
    + J_2 \sum_{\langle\langle i, j \rangle\rangle} \mathbf{S}_i \cdot \mathbf{S}_{j} \\
    &+ J_{\mathrm{Z}} \sum_{\langle i, j \rangle} S^z_i S^z_j
    + J_{3} \sum_{\langle\langle\langle i, j \rangle\rangle\rangle} \mathbf{S}_i \cdot \mathbf{S}_{j} \\
    &+ J_{\mathrm{DMI}} \sum_{\langle i, j \rangle} \mathbf{D} \cdot [\mathbf{S}_i \times \mathbf{S}_j]
\end{split}
\end{equation}
Where $J_1$, $J_2$, $J_3$, $J_{\mathrm{Z}}$, and $J_{\mathrm{DMI}}$ are the nearest, next-nearest, second-next-nearest, nearest anisotropic, and antisymmetric Dzyaloshinskii-Moriya interaction (DMI) spin exchanges. As the spin models are on top of a surface, we take $\mathbf{D} = (0,0,1)$. Placing a spin chain on a substrate breaks mirror symmetry, and together with spin-orbit coupling causes an antisymmetric DMI to occur~\cite{Choi2019, Bode2007, Lee2015}. Strong spin-orbit coupling combined with superexchange leads to anisotropic exchange~\cite{Johnson2015}. In light elements, anisotropic exchange and DMI are typically much smaller than the corresponding isotropic exchange interactions~\cite{Camley2023}. In heavier elements, isotropic and anisotropic interactions can become comparable,
and can be tuned by choosing suitable elements and adjusting their mutual distances~\cite{Khajetoorians2016}. Together with the substrate engineering ~\cite{Camley2023, Kuepferling2023}, all the interactions can be driven to obey the same magnitude.

Our objective is to use the ability of STM to measure spin excitations with spatial and frequency resolution, to learn the many-body Hamiltonian of a quantum magnet.
For a pristine system, as the length of the spin chain increases, the many-body excitations on different spin sites start to become identical
due to the disappearing finite-size effects. As a result, the spatial resolution of STM
would not provide additional information in the limit of pristine infinite chains.
In contrast, for finite spin chains, confined many-body spin modes will appear in the system,
which directly reflect the dispersion of the many-body excitations.
In general, introducing local impurities creates
site-dependent excitations around the impurity, information that our machine
learning methodology will leverage to learn the Hamiltonian of the system. 

The role of impurities in our approach goes beyond simply breaking translational symmetry. 
While pristine finite chains host confined modes, their dynamical correlators become increasingly homogeneous with system size, and different Hamiltonian terms can produce highly correlated spectroscopic signatures, making parameter disentanglement fragile, especially in the presence of noise. 
Local impurities act as controlled perturbations that selectively couple to many-body eigenstates, inducing spatially resolved reconstructions of the excitation spectrum that are strongly sensitive to the microscopic Hamiltonian. 
This effect is further enhanced when multiple impurity configurations are considered, as varying impurity number and separation provides complementary real-space responses that improve parameter identifiability and robustness.

In the following, we will focus on moderately large quantum spin models
with one or several impurities, whose spectra show a complex
interplay between confined modes and impurity reconstructions.
Localized impurities can be engineered by depositing additional atoms close to
the spin chain. For the sake of concreteness, we take that the
impurity atom only couples to the site closest to it as: 
\begin{equation}
\begin{split}
    \label{perturbation}
    H_{\mathrm{impurity-chain}} &= \lambda \sum_{\langle \alpha_i, \beta_j \rangle} (\mathbf{S}_{\alpha_i} \cdot \mathbf{S}_{\beta_j})
\end{split}
\end{equation}
where $\lambda$ is the strength of the perturbation, $\alpha_i$ is the spin site in the chain to which the closest impurity couples, and $\beta_j$ is the site corresponding to the impurity atom.

\subsection{Spin spectral function}
\label{Spin spectral function and tunneling junction}

The spin Hamiltonian determines the different energies at which spin excitations can occur in the system. The dynamical correlator provides information on all possible excitations at zero temperature:
\begin{equation}
\label{eqn: dynamical correlator}
    S^{aa}_n(\omega) = \langle \mathrm{GS} | S_n^a \, \delta(\omega - H + E_{\mathrm{GS}}) \, S_n^a | \mathrm{GS} \rangle,
\end{equation}
where $S_n^a$ is the spin operator $\mathbf{S}$ acting on the site $n$ for which $a \in \{x, y, z\}$, $|\mathrm{GS}\rangle$ is the ground state of the many-body Hamiltonian, $\omega$ is the frequency, and $E_{\mathrm{GS}}$ is the ground-state energy.

The spectral function above can be explicitly expanded by inserting a complete set of eigenstates $|\alpha\rangle$ of the Hamiltonian, which makes clear that the correlator encodes all possible spin excitations:
\begin{equation}
\delta(\omega - H + E_{\mathrm{GS}}) = \sum_\alpha |\alpha\rangle \langle \alpha| \, \delta(\omega - E_\alpha + E_{\mathrm{GS}}),
\end{equation}
so that
\begin{equation}
S^{aa}_n(\omega) = \sum_\alpha |\langle \alpha | S^a_n | \mathrm{GS} \rangle|^2 \, \delta(\omega - E_\alpha + E_{\mathrm{GS}}).
\label{eq:Saa_expanded}
\end{equation}
This formulation shows that the dynamical correlator $S^{aa}_n(\omega)$ directly contains the spectral weight of the excited states $\alpha$, each corresponding to one additional spin excitation relative to the ground state. This object can be computed with a tensor network Chebyshev kernel polynomial~\cite{PhysRevB.83.195115,RevModPhys.78.275,Lado2019,PhysRevResearch.2.023347,10.21468/SciPostPhysCodeb.4,dmrgpy}.

\begin{figure*}[t!]
    \centering
    \includegraphics[width=0.9\linewidth]{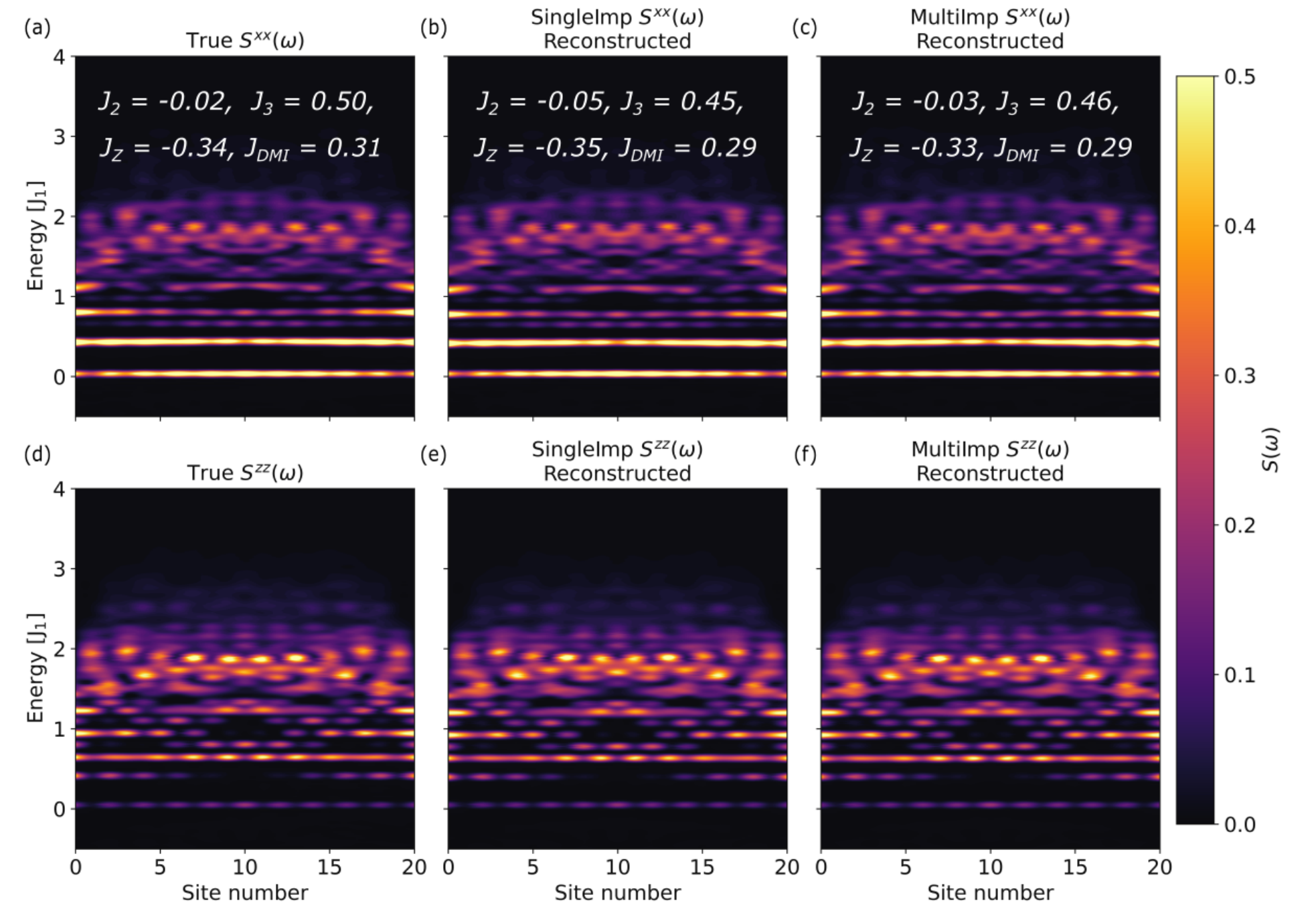}
    \caption{\textbf{Noiseless Hamiltonian learning}.
    Comparison between true dynamical correlators (a,d),
    and dynamical correlators obtained from the parameters by the Hamiltonian
    learning algorithms (b,c,e,f), using the SingleImp (b,e) and MultiImp (c,f) algorithms.
    Panels (a,b,c) show the $S_{xx}$ dynamical correlator, and $d,e,f$ the $S_{zz}$ dynamical correlator.
    Predictions are made under no-noise conditions of $\chi=0.0$ and with impurity coupling of $\lambda=0.122$. We have observed
    that for pristine dynamical correlators, both networks perform similarly.
}
    \label{fig:reconstructed dcs chi = 0} 
\end{figure*}

\begin{figure*}[t!]
    \centering
    \includegraphics[width=0.9\linewidth]{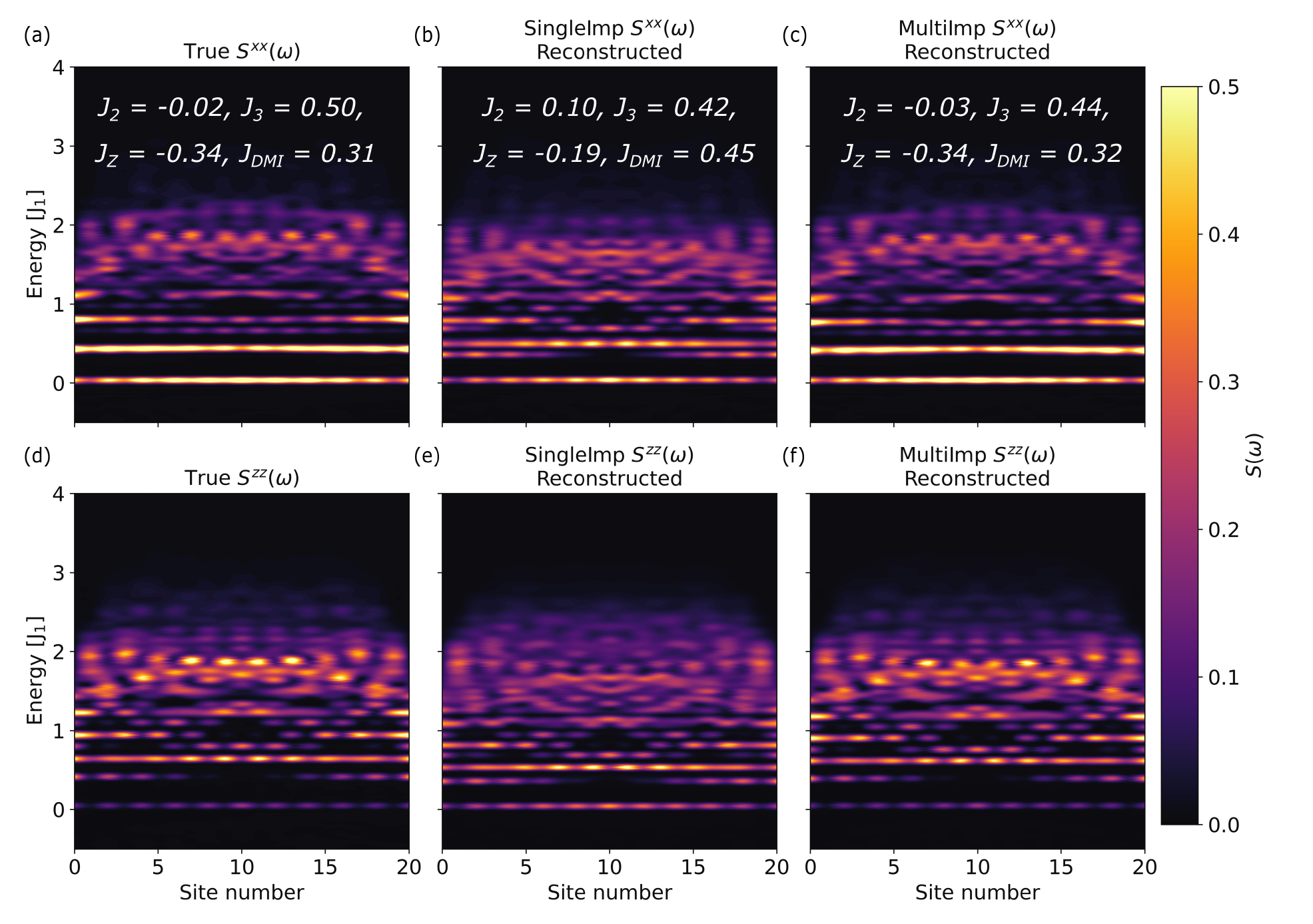}
    \caption{
    \textbf{Noisy Hamiltonian learning}. Many-body dynamical correlators, comparing true (a,d) and predicted many-body excitations (b,c,e,f),
    obtained by SingleImp (b,e) and MultiImp (c,f) algorithms. Predictions are made under noise conditions of $\chi=1.0$ and with impurity coupling of $\lambda=0.122$.
        Panels (a,b,c) show the $S_{xx}$ dynamical correlator, and $d,e,f$ the $S_{zz}$ dynamical correlator.
        It is observed that the MultiImp network performs a more faithful prediction,
    in particular, leading to spectral functions that agree with the original ones in small features
    that SingleImp does not account for.}
    \label{fig:reconstructed dcs chi = 1} 
\end{figure*}

These spin excitations can be probed with STM-electron spin resonance (ESR)~\cite{Yang2019, Baumann2015, Paul2016, Phark2023, Wang2023, Lado2017, Zhao2025, Seifert2021} or via inelastic electron tunneling spectroscopy (IETS)~\cite{Heinrich2004, Kgel2018, Gawronski2008, Lorente2009, Lorente2005}. 
In tunneling spectroscopy,
the dynamical correlator in eq.~\ref{eqn: dynamical correlator} corresponds to the second derivative of the measured tunneling current through the sample:
\begin{equation}
\label{eqn: dynamical correlator approx}
S^{aa}_n(\omega) \sim \frac{\mathrm{d^2}I}{\mathrm{d}V^2}
\end{equation}
where $I$ is the observed dc tunnel current through the sample, $V$ is the applied bias voltage, and $\omega$ is the frequency~\cite{Paul2016}.

We note that this relation holds at low energies, well below the
electronic repulsion of the sites.
The inferred spin Hamiltonian should be interpreted as an effective low-energy model capturing the dominant interactions governing the observed excitations.
Furthermore, the direct relationship of
Eq.\ref{eqn: dynamical correlator approx} 
holds at zero temperature, and in the absence of Kondo coupling and
out-of-equilibrium effects. 
However, in the presence of Kondo coupling\cite{Otte2008,Ternes2015} or
out-of-equilibrium dynamics\cite{Loth2010,PhysRevB.82.134414}, such direct correspondence is no longer valid,
and a full many-body solution of the current is required.
It is worth noting that while we do not perform such generalized solution in our manuscript, our methodology can be adapted to those scenarios by replacing
our computed dynamical correlator by the many-body solution of the $dI/dV$\cite{Ternes2015,PhysRevB.82.134414}.

In the following, we perform the training of the algorithm
with the cumulative integrals of the dynamical correlators, 
equivalent to the differential conductance in spectroscopy
\begin{equation}
    \int_0^V S(\omega) \textit{d}\omega \sim \frac{\mathrm{d}I}{\mathrm{d}V}
\end{equation}

The many-body spin excitations depend on the Hamiltonian, which is directly reflected in the dynamical correlators. For Hamiltonians
with a single dominant parameter, 
the dynamical excitations are easily distinguishable.
In contrast, 
quantitatively extracting the value 
of multiple parameters is
a remarkable challenge,
especially when several 
fall within comparable magnitudes and no single parameter is dominant. Distinguishing these correlators directly
represents a nontrivial problem, as the isotropic terms $J_2$ and $J_3$ produce similar excitations, and the anisotropic terms also comparably influence the correlators. Figure \ref{Fig:schematic} (b) shows how different dominating parameters affect the appearance of the dynamical correlators. Furthermore, experiments include noise that affects the available data,
 making parameter extraction even more challenging. 

\section{Machine learning methodology}
\label{section: hamiltonian learning}

Mapping the dynamical correlators to the underlying Hamiltonian cannot be achieved in a straightforward manner. Here, we present an approach to address this complex inverse problem by using computationally generated dynamical correlators to guide parameter inference through machine learning. Since the bias-integrated dynamical correlator directly corresponds to the differential conductance $\frac{dI}{dV}$ measured in STM experiments, this establishes a clear and quantitative bridge between theory and experiment. In this way, our method enables an experimentally realistic analysis of spin excitations, allowing the extraction of Hamiltonian parameters and providing valuable insight into the fundamental properties of the system.

Here, we employ a supervised neural network (NN) to infer Hamiltonian parameters from the dynamical correlators of spin chains. We note that the model is implemented as a feedforward neural network (FFNN), which is well suited to our principal component analysis (PCA) compressed spectral inputs where relevant information is encoded in nonlocal correlations rather than local features. The details of the NN architecture and training procedure are provided in Appendix~\ref{appendix: neural network details}. 

\subsection{Impurity configurations in spin chains}
\label{selected systems}

We consider a spin chain consisting of 21 spins and analyze three different systems, each characterized by distinct impurity configurations obtained by tuning the separation between the impurities (Fig.~\ref{Fig:schematic}). To motivate our choice of systems, we note that in an infinitely long spin chain the physics is translationally invariant. Therefore, shifting a single impurity from one site to another does not alter the system. In a finite chain of 21 spins, we instead vary the number and symmetric placement of impurities.

In our setup, the spin chain sites are numbered starting from zero. For the first system, we placed a single-impurity atom adjacent to the middle of the chain, at site~10. In the second configuration, two impurities were positioned adjacent to sites~9 and~11, while in the third configuration, two impurities were placed adjacent to sites~8 and~12, three sites apart from each other. This progression allows us to mimic the infinite-chain behavior while preserving reflection symmetry and minimizing boundary effects, ensuring that any differences in the results can be attributed to impurity number and separation rather than trivial edge effects.

\subsection{Data generation}
\label{data generation}
We computed more than 400 dynamical correlators $S(\omega)$ for the $xx$ and $zz$ components across the different systems using the MPS-KPM method. The $zz$ component was included because the impurity spins are primarily coupled along the $z$ axis. Since the Dzyaloshinskii-Moriya interaction with $\mathbf{D} \parallel z$ couples the $x$ and $y$ spin components symmetrically, we included only the $xx$ component to represent the transverse spin response.

The Hamiltonian parameters were randomly sampled from a uniform distribution for each sample and kept identical across all three systems. Details of data generation and parameter ranges can be found in the Appendix \ref{appendix: data}. For each system, we took the cumulative integrals of the correlators, and the resulting samples were concatenated to form the dataset. Two datasets were used for training. The first dataset included only the correlators where the impurity atom was adjacent to the middle site (Fig \ref{Fig:SingleImp and MultiImp} (a)) and it was used to train the SingleImp network. The second dataset was constructed by concatenating correlators from all three impurity placements (Fig \ref{Fig:SingleImp and MultiImp} (b)), ensuring that correlators with identical Hamiltonian parameters were combined into a single sample. This dataset was used to train the MultiImp network. 

\subsection{Inclusion of noise}
\label{subsection: noise in dynamical correlators}

To demonstrate that our algorithm adopts an experimentally realistic approach to Hamiltonian parameter learning, we incorporate controlled stochastic noise into the tunneling junction signal. The noise represents random experimental fluctuations and is modeled as an additive, frequency-dependent Gaussian contribution:
\begin{equation}
\left(\frac{dI}{dV}\right)_{\mathrm{noisy}}(\omega) = \frac{dI}{dV}(\omega) + \zeta(\omega),
\end{equation}
where $\zeta(\omega)$ denotes a Gaussian random variable with zero mean and width $\chi \cdot \sigma$. Here, $\sigma$ is the standard deviation of the $\frac{dI}{dV}(\omega)$ values across the training or test set, and $\chi$ is a noise-strength parameter controlling the overall level of fluctuations. 

We note that the noise parameter $\chi$ is introduced as a normalized, dimensionless quantity that scales the noise amplitude relative to the intrinsic variance of the tunneling spectra, rather than being calibrated to a specific experimental setup.
In practice, $\chi$ acts as phenomenological control parameter that scales the noise amplitude relative to the intrinsic variance of the tunneling spectra, 
including in an effective manner noise arising from
multiple different sources~\cite{PhysRevResearch.4.033223,lupi2024,2024arXiv240504596V,PhysRevApplied.20.044081,Lupi2026,Koch2025,Karjalainen2023,Nigmatulin2026}.
This choice enables a system independent benchmarking of robustness, avoiding dependence on setup-specific quantities such as absolute current levels or tip conditions. 
Noise levels $\chi \sim 0.5$--$1$ correspond to regimes where spectral features are visibly broadened but remain experimentally resolvable, consistent with typical STM-IETS and STM-ESR measurements~\cite{Hwang2022,Reed2008}.

For each Hamiltonian realization, $\zeta(\omega)$ is independently sampled at every discrete frequency point, resulting in uncorrelated Gaussian offsets along the spectrum. Further implementation details are provided in Appendix~\ref{appendix: neural network details}.

\subsection{Robustness against noise}
To evaluate how the prediction accuracy decays as a function of noise strength, we compute the fidelities of the predictions
defined as the following metrics~\cite{Khosravian2024,lupi2024,Koch2025}:
\begin{equation}
    \label{eqn:fidelity}
    \mathcal{F} (\Lambda^{\text{pred}}, \Lambda^{\text{true}}) = \frac{|\langle \Lambda^{\text{pred}} \Lambda^{\text{true}} \rangle - \langle \Lambda^{\text{pred}} \rangle \langle \Lambda^{\text{true}} \rangle|}{\sqrt{\mathcal{C}(\Lambda^{\text{true}})\mathcal{C}(\Lambda^{\text{pred}})}}
\end{equation}
where $\Lambda^{\text{true}}$ represents the true values, $\Lambda^{\text{pred}}$ represents the predicted values, 
$\mathcal{C}(X) = \langle X^2 \rangle - \langle X \rangle^2$
and $\mathcal{C}(X) = \langle X^2 \rangle - \langle X \rangle^2$
denotes the cumulant of a certain variable $X$. The fidelity $\mathcal{F}$ just takes values in the $[0, 1]$, where $\mathcal{F} = 1$ indicates perfect prediction accuracy ($\Lambda^{\text{true}}=\Lambda^{\text{pred}}$) while $\mathcal{F} = 0$ corresponds to no predictive accuracy.

\begin{figure}[t!]
\includegraphics[width=\columnwidth]{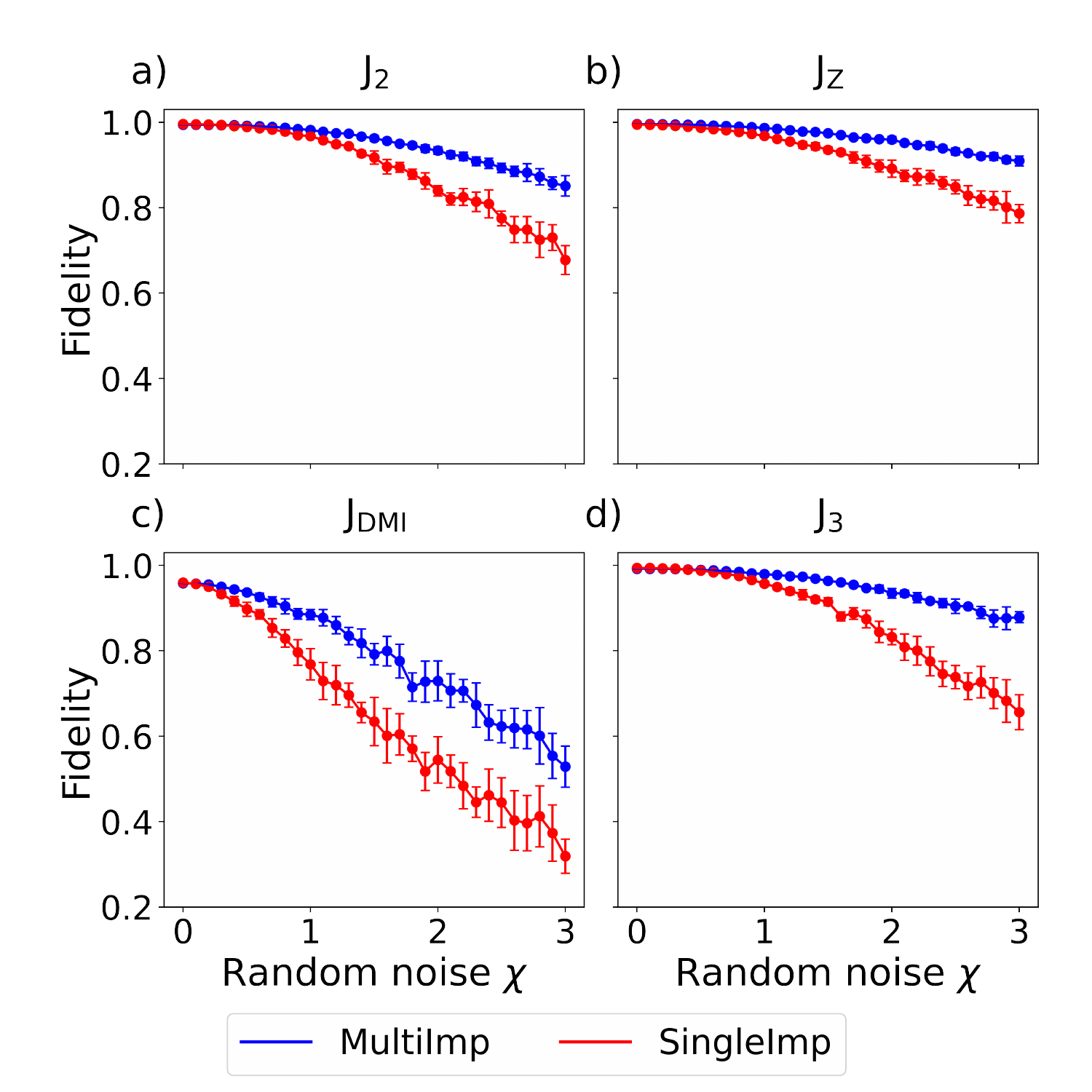}
\caption{\textbf{Fidelities of Hamiltonian learning as a function of increasing amount of noise $\chi$.} The NN SingleImp is trained with single impurity dataset Fig \ref{Fig:SingleImp and MultiImp} (a) and the MultiImp is trained with concatenated datasets from multiple impurity placements Fig \ref{Fig:SingleImp and MultiImp} (b). Error bars indicate the standard deviation of fidelity values obtained over ten stochastic runs.}
\label{Fig: fidelities_subplots}
\end{figure}

\section{Noiseless and noisy Hamiltonian learning}
\label{section: discussion}
We first exemplify our algorithm with non-noisy data, demonstrated in Fig \ref{Fig: fidelities_subplots}.
While the limit of noiseless data is not experimentally significant,
it provides a useful demonstration of the single impurity and multi-impurity strategies.
Both networks perform similarly with perfect data, having the fidelity values of $\mathcal{F}_{\mathrm{J2}}=0.99, \mathcal{F}_{\mathrm{JZ}}=0.99, \mathcal{F}_{\mathrm{J3}}=0.99$, and $\mathcal{F}_{\mathrm{JDMI}}=0.96$. This is illustrated in the Fig \ref{fig:reconstructed dcs chi = 0}, where dynamical correlators are reconstructed via DMRG-KPM based on the predictions provided by both networks. The predictions of $J_\mathrm{{DMI}}$ are slightly worse compared to the other parameter predictions.
The lower accuracy in the prediction of $J_\mathrm{{DMI}}$
can be rationalized from the fact that extracting effects stemming
from off-diagonal coupling is a more challenging problem. This will
be observed even more clearly in the presence of noise.

To evaluate the robustness of our algorithm, 
we now use the trained networks to predict the parameters from the noisy data under varying levels of input noise. The fidelity between the predicted and true values was calculated for each parameter of interest. This process was done across 10 stochastic runs for each noise level to obtain a distribution of fidelities. For each value of the noise parameter $\chi$, Gaussian noise with zero mean and standard deviation proportional to $\chi$ was added to the test data. We used unseeded randomness to simulate different noise realizations in each run.
The average fidelities for each value of $\chi$ were computed, and the results are shown
in Fig.~\ref{Fig: fidelities_subplots}. With noise of $\chi = 1.0$, the fidelities for the single-impurity algorithm SingleImp are 
$\mathcal{F}_{\mathrm{J2}}=0.96, \mathcal{F}_{\mathrm{JZ}}=0.97, \mathcal{F}_{\mathrm{J3}}=0.96$, and $\mathcal{F}_{\mathrm{JDMI}}=0.82$, whereas for MultiImp fidelities are
$\mathcal{F}_{\mathrm{J2}}=0.98, \mathcal{F}_{\mathrm{JZ}}=0.98, \mathcal{F}_{\mathrm{J3}}=0.98$, and $\mathcal{F}_{\mathrm{JDMI}}=0.89$. 
When evaluating with noisy dataset, MultiImp performs significantly better than SingleImp.  The fidelities for $J_2$, $J_{\mathrm{Z}}$, and $J_3$ 
remain almost stable even up to $\chi=1.5$, whereas the fidelity for $J_\mathrm{{DMI}}$ decreases remarkably when $\chi > 1.0$. It is worth noting that fidelity values remain higher for all parameters when the network is trained with concatenated datasets from multiple impurity placements. This is demonstrated in Fig \ref{fig:reconstructed dcs chi = 1}. By extracting the Hamiltonian parameters via both neural networks and reconstructing the dynamical correlators via DMRG-KPM, the MultiImp reconstructed dynamical correlators correspond better to the true correlators than the SingleImp reconstructions.

The lower fidelity in the prediction of $J_{\mathrm{DMI}}$ can be attributed to the fact that the network is trained using only diagonal dynamical correlators $S^{aa}(\omega)$, while neglecting off-diagonal terms $S^{ab}(\omega)$ with $a \neq b$, which more directly encode antisymmetric interactions. 
Including such contributions would likely improve the accuracy of $J_{\mathrm{DMI}}$, but they remain challenging to access experimentally, providing a practical justification for our approach.

We also emphasize that the Hamiltonian considered here does not aim to exhaustively capture all interactions present in real quantum magnets. 
Instead, we focus on the dominant exchange, anisotropic, and Dzyaloshinskii-Moriya terms relevant for a broad class of spin-$1/2$ systems on surfaces. 
Additional effects, such as further-neighbor couplings, impurity-induced renormalizations, or dipolar interactions, may be present but are typically subleading in experimentally relevant regimes. 
Importantly, the proposed framework is not restricted to this specific model: Additional interaction terms can be incorporated by extending the training dataset and output parameter space. 
In this context, multi-impurity tomography is particularly advantageous, as it provides complementary information that enhances parameter identifiability when multiple interactions compete. 
Overall, our results establish a robust and experimentally relevant proof of principle for Hamiltonian learning, rather than a complete microscopic description of all possible systems.

Finally, we note that our methodology assumes that impurities act as weak, local perturbations probing a fixed underlying many-body Hamiltonian. 
As such, it is directly applicable to localized spin systems dominated by short-range exchange interactions. 
Regimes involving strong hybridization with itinerant electrons, Kondo screening, or impurity-induced modifications of the effective Hamiltonian fall outside the scope of the present work. 
In such cases, alternative approaches based on effective low-energy models or data-driven methods tailored to itinerant systems may be required. 

\section{Conclusion}
\label{section: conclusions}
Here, we demonstrated a machine learning strategy that leverages local impurity
spins to extract the spin Hamiltonian of quantum magnets.  Our methodology
relies on using the reconstruction of the many-body spin excitations to map the
underlying Hamiltonian by exploiting the frequency and spatial resolution of
scanning probe spectroscopy.  Our methodology enables learning of complex
quantum many-body Hamiltonians, which is often challenging to achieve using
traditional techniques.  Our results are based on simulated dynamical
correlator functions, which enabled training a machine learning algorithm.
Using dynamical correlators that are available in IETS and STM-ESR
measurements, the underlying parameters can be extracted from measurements
reflecting local spin excitations.  We demonstrated two algorithms, one using a
single impurity and a second one using multiple impurities simultaneously.  The
usage of several impurities provides robustness against significant noise in
the dynamical spin excitations, a feature of substantial importance for
realistic Hamiltonian inference from experimental data.  Furthermore, while
Hamiltonian extraction could be performed from a multidimensional fitting, such
a procedure requires solving a sequential set of Hamiltonians, which becomes
unfeasible for complex quantum many-body Hamiltonians.  In stark contrast, our
methodology allows us to extract the parameters from the measured spectral
function instantly once it is trained.  Finally, it is worth noting that
although our strategy focused on quantum spin models, analogous methodologies
can enable learning Hamiltonians of other complex strongly correlated states of
matter, ranging from correlated metals, fractional topological states, and
correlated superconductors.  Our methodology puts forward impurity-driven
excitations as a flexible knob to train machine learning methodologies to
perform Hamiltonian learning in quantum many-body magnets.

\textbf{Acknowledgements}:
We acknowledge financial support
from InstituteQ, the Finnish Quantum Flagship, 
the ERC Consolidator Grant ULTRATWISTROICS (Grant agreement no.
101170477),
the
Research Council of Finland (RCF Research Fellow no.~369367, 
and RCF project no. 370912), and 
the Finnish Centre of Excellence in Quantum Materials QMAT (No. 374166).
We acknowledge the computational resources provided by the Aalto Science-IT project.
R.K. acknowledges support from the IGNITE project under grant agreement no.~101069515 of the Horizon 
Europe Framework Programme and the KIND synergy program from the Kavli Institute of Nanoscience Delft.
We thank R. Drost and P. Liljeroth for useful discussions.

\section*{Data availability}
The data that support the findings of this article are openly available in Refs.~\cite{nettacode,nettazenodo}.

\appendix

\section{Many-body data generation}
\label{appendix: data}
We took $J_1$ as the energy scale and randomly sampled the parameters $J_2, J_{\mathrm{Z}}$ and $J_3$ in the range $\in[-0.5, 0.5]$ and $J_{\mathrm{DMI}} \in [0, 0.5]$. $J_{\mathrm{DMI}}$ values were drawn from the positive range, since the sign of the interaction does not produce a difference in the dynamical correlators. We also sampled the perturbation strength randomly from an interval $\lambda \in [0.1, 0.3]$ for each sample. In this range, the perturbation is great enough to produce a significant effect, but small enough to remain as a perturbation to the system. The parameter ranges were chosen to reflect the range of interactions observed in the experimental setups and be suitable for the training of the machine learning model. These intervals also ensured that the Hamiltonian captures both ferromagnetic and antiferromagnetic couplings, as well as possibly other exotic quantum phases. 

\section{Neural network architecture}
\label{appendix: neural network details}
The Hamiltonian parameters were normalized to the range $[0,1]$ before training of the networks. We computed ten different noisy copies of each data sample with increasing values of $\chi$ up to $\chi = 0.1$, where the noise width was defined as $\chi \cdot \sigma$ and $\sigma$ is the standard deviation of the $\frac{dI}{dV}(\omega)$ values across the training set. The neural network was then trained using both the noiseless dataset and the noisy samples to improve accuracy under experimentally realistic conditions.

We used the network architecture shown in table \ref{tab:nn architecture}.

Each layer uses ReLU as an activation function and kernel initializer GlorotUniform(3) for repeatability. A dropout layer is used as a regularization method with a dropout percentage $5 \%$. We apply PCA on the dataset with the cumulative explained variance of $99\%$ to reduce noise and dimensions of the dynamical correlators. We chose ADAM as an optimizer and ran the NN over 600 epochs with a batch size of 100. We evaluated the model by observing mean squared error loss and fidelity.
\begin{table}[h!]
    \centering
    \begin{tabular}{l|c}
    Layer type & nodes \\
    \hline
    InputLayer & num of PCA components\\
    Dense & 500 \\
    Dropout & 5 $\%$ drop-off\\
    Dense & 200 \\
    Dense & 100 \\
    Dense & 4
    \end{tabular}
    \caption{NN architecture. The number of PCA components: 124 for MultiImp and 126 for SingleImp.}
    \label{tab:nn architecture}
\end{table}

To compute the fidelities, we introduced random noise into the dynamical correlator testing dataset. For each noise level $\chi$, we generated ten independent noisy datasets by adding different realizations of random noise to the correlators, where the noise width was defined as $\chi \cdot \sigma$ and $\sigma$ corresponds to the standard deviation of the $\frac{dI}{dV}(\omega)$ values across the test set. The trained models were then used to predict the system parameters for each noisy dataset, and the fidelities of these predictions were averaged. This process was repeated for a range of increasing $\chi$ values, allowing us to obtain averaged fidelities (Fig.~\ref{Fig: fidelities_subplots}) and smooth curves despite the inherent randomness introduced by the noise.

It is worth noting that we employ an FFNN rather than a convolutional architecture. 
This choice is motivated by the structure of the input data: After PCA preprocessing, the dynamical correlators are represented as global feature vectors where relevant information is distributed across nonlocal correlations. 
In this setting, convolutional filters designed to capture local patterns do not provide a clear advantage. 
We emphasize that the methodology is not restricted to this architecture, and alternative models could be explored for different data representations.

The trained models and data are available in GitHub\cite{nettacode} and Zenodo\cite{nettazenodo}.

\begin{figure}
    \centering
    \includegraphics[width=\linewidth]{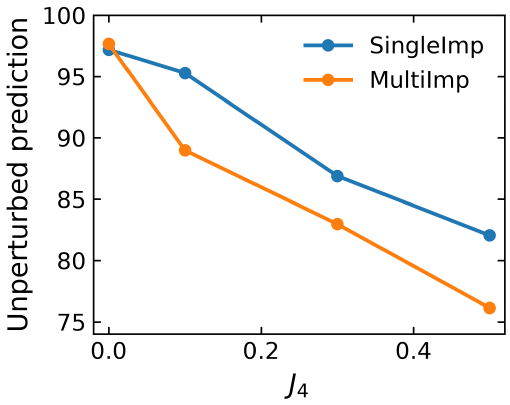}
    \caption{
    Prediction accuracy $\left \langle \Xi(J_n)\right \rangle _{n\le 3}$ vs the strength of an unmodeled fourth-neighbor interaction $J_4$. The models are trained without $J_4$ and tested on spectra generated with finite $J_4$. Accuracy is computed from Eq.~\ref{eq:acc_main} and averaged over all exchange couplings.
}
    \label{fig:j4_misspecification}
\end{figure}
\section{Robustness against model misspecification}
\label{app:model_misspecification}
One possibility in experiments is that a certain interaction is present
that was not originally considered in the theoretical training set,
which is known as model misspecification.
To assess the robustness of the Hamiltonian-learning framework against model misspecification, we perform a controlled validation test in which the neural network is trained on the Hamiltonian introduced in Sec.~\ref{section: quantum spin chains}, while the test spectra are generated from an extended Hamiltonian including an additional fourth-neighbor exchange interaction $J_4$ not present during training.

Starting from the representative sample shown in Fig.~\ref{fig:reconstructed dcs chi = 0}, we generate spectra for increasing values of $J_4$ and evaluate the reconstruction accuracy of both the SingleImp and MultiImp models. For each exchange coupling, we define the unperturbed prediction ($\Xi$) as
\begin{equation}
\Xi(J_n)=100\left(1-\frac{|J_n^{\mathrm{pred}}-J_n^{\mathrm{true}}|}{J_n^{\max}-J_n^{\min}}\right),
\label{eq:acc_main}
\end{equation}
where $J_n^{\max}$ and $J_n^{\min}$ correspond to the limits of the training range. The reported accuracy is obtained by averaging over all exchange couplings of the test sample.

Figure~\ref{fig:j4_misspecification} shows the resulting unperturbed prediction as a function of $J_4$. Both models achieve high accuracy in the absence of model mismatch ($J_4=0$), while the reconstruction quality progressively decreases as the strength of the unmodeled interaction increases. This behavior is expected, since the spectral signatures associated with $J_4$ may not be represented within the Hamiltonian class used during training and are therefore absorbed into the inferred shorter-range couplings.
The current calculation shows the ability of the algorithm to represent spectra without $J_4$, but itself is not a measure of accuracy for non-zero $J_4$.

These results indicate that the method is robust against weak deviations from the assumed Hamiltonian, while providing a clear quantitative signature when relevant interaction terms are missing from the model. In principle, such missing interactions can be incorporated straightforwardly by extending the training Hamiltonian and treating the corresponding coupling constants as additional learning targets. For example, in the present case, the reconstruction of spectra containing finite $J_4$ interactions would require generating training data with varying $J_4$ values and including $J_4$ among the parameters inferred by the neural network.

\bibliography{references.bib}

\end{document}